\documentstyle[pra,aps]{revtex}

\newcommand{\beq}{\begin{equation}} 
\newcommand{\eeq}{\end{equation}} 
\newcommand{\lab}{\label}

\title{THE PROBLEM  OF  TIME  IN  PARAMETRIZED THEORIES\footnote{Published in
General Relativity and Gravitation {\bf 26}, 1267-1275 (1994)}}

\author{Fabi\'an H.  GAIOLI and  Edgardo  T.  GARCIA-ALVAREZ\footnote{E-mail:
gaioli@iafe.uba.ar;  \ galvarez@dfuba.df.uba.ar}}  

\address{Departamento  de  F\'{\i}sica,    Facultad  de  Ciencias  Exactas  y
Naturales, \\ Universidad de Buenos Aires, 1428 Buenos Aires, Argentina \\ \&
\\ Instituto de Astronom\'\i a  y  F\'\i  sica del Espacio, \\ C.C.  67, Suc.
28, 1428 Buenos Aires, Argentina} 

\date{\today}

\vspace{1.0in}

\begin{document}
\maketitle

\begin{abstract}

A  common  feature  of reparametrization invariant theories is the difficulty
involved  in    identifying    an  appropriate  evolution  parameter  and  in
constructing a Hilbert  space  on  states.    Two well known examples of such
theories are the relativistic point particle and the canonical formulation of
quantum  gravity.    The  strong    analogy    between  them  (specially  for
minisuperspace models) is considered in order  to  stress  the correspondence
between  the  ``localization  problem''  and  the  ``problem    of    time,''
respectively.  A possible solution for the first  problem  was  given  by the
proper  time  formulation  of  relativistic  quantum  mechanics.    Thus,  we
extrapolate  the  main  outlines of such a formalism to the  quantum  gravity
framework.  As a consequence, a proposal to solve the problem of time arises. 

\end{abstract}

\vskip 2cm Pacs numbers: 03.65.Pm, 04.60.+n  

\vskip 15cm 

\narrowtext
\twocolumn

Since  B.S.    DeWitt's  celebrated  paper  \cite{dw1}  about  the  canonical
formulation of quantum gravity (QG)  was  published,  the problem of time and
the interpretation of the wave functional have remained unsolved. 

The main difficulty in  reconciling  the  antagonistic  formalisms of quantum
mechanics and the theory of relativity lies in the role and interpretation of
``time.'' This is true not only  in  the case of general relativity, but also
at the level of the special theory  of relativity.  In fact, while the latter
theory  deals with the space-time coordinates on an  equal  footing,  quantum
mechanics   privileges  an  external  absolute  parameter  which  labels  the
evolution  of  the  state  of  the system.  Therefore, in  the  second  case,
``time'' should have the properties of a $c$-number, unlike in the first case
where,  since  the spatial coordinates are raised to the status of operators,
Lorentz  transformations  should  impose   this  character  on  the  temporal
coordinate as well.  Thus, this dual role of ``time'' generates a conflict in
relativistic quantum mechanics (RQM).  This  is  basically  the  reason why a
one-particle theory in the usual formulation of  RQM  has  several conceptual
difficulties,  such  as  the interpretation of negative energies  which  gave
origin  to the famous ``problem of localization.'' (An exhaustive  review  of
the localization problem up to the 70s can be found in \cite{kal}). 

The  history  of  physics  teaches  us that every conflict arising  from  the
unification of  two  fundamental  theories  reveals the incompatibility among
hypotheses of the  two  theories.    Thus,  some  of them must be removed and
therefore some prejudices we firmly cling to must be overcome.  

In the 40s a  proper  time  formalism  (PTF) was developed \cite{fsfns} for a
possible unification between special relativity and quantum mechanics, within
the framework of a consistent one-particle  theory.   However, the price that
we must pay to celebrate this wedding  is  to give up the concept of definite
mass state. 

The PTF introduces an evolution absolute parameter $\tau$, independent of the
proper time but related to it through the classical  limit \cite{pla1}, which
parametrizes the dynamics of the system.  This parameter plays  the  role  of
``time'' \cite{hor,son},  while  the  temporal  coordinate  has  a  different
status,  being promoted  to  the  rank  of  operator.    The  classical  mass
constraint  is  removed  but  nevertheless  the  standard  RQM  is  recovered
on-shell.  In this way  the  formalism closely copies the general outlines of
the  nonrelativistic quantum mechanics, furnishing the  theory  with  a  well
known structure. 

This  matter  has recently gained renewed interest  with  the  appearance  of
several works that intended to go beyond the  formal  aspects.   (Versions of
RQM similar to the one presented here have been  considered,  e.g.,  in Refs.
\cite{fsfns,hor,son}.  See also Refs.  \cite{todos,jon}). 

In the last decade the analogy between the canonical formulation  of  QG  and
the  spinless    relativistic   point  particle  has  been  widely  exploited
\cite{cos}.  Therefore,  it  is  not  surprising  to  find  the  same kind of
difficulties since both systems  are  similar  in  nature.   These issues are
known in QG as ``the problem of time'' \cite{kuis}.   

The aim of this essay  is  two-fold.    On the one hand we try to connect two
lines of research, the localization problem in RQM and the problem of time in
QG.  On the other hand, we propose a parametrized theory for QG, which can be
considered as a promising solution to the problem of time. 

For the sake of simplicity, in this essay  we  only  consider  minisuperspace
models for QG \cite{hal1}.  The full superspace formalism  will  be discussed
elsewhere \cite{cgga}.

Let us start with the spinless relativistic point particle in  the  PTF.  Let
$\{\vert  x^{\mu}\rangle\}$ ($\mu =0,1,2,3$) be the basis of localized states
of the  position  operator  $X^{\mu}$  spanning a linear space endowed with a
scalar product,

\beq
\langle\Phi\vert\Psi\rangle  =    \int    d^4x  \Phi^*(x^{\mu})\Psi(x^{\mu}),
\lab{pe}
\eeq

\noindent
satisfying the normalization and the completeness conditions,  

\beq
\langle x^{\mu}\vert x'^{\mu}\rangle =\delta (x^{\mu} - x'^{\mu}), \  \ \ \ \
\int d^4x \vert x^{\mu}\rangle\langle x^{\mu}\vert = {\rm 1\!\!I}. 
\lab{nor}
\eeq

\noindent
In this coordinate  representation  the state of the system is represented by
the wave function, belonging to a four-dimensional Hilbert space,   

\beq
\Psi(x^{\mu})=\langle x^{\mu}\vert\Psi\rangle, \lab{fo}
\eeq

\noindent
defined on the space-time manifold.   The position operator and its canonical
conjugate variable, the momentum $P_{\mu}$, satisfying ($\hbar = c = 1$)

\beq
[X^{\mu},  P^{\nu}]  =  -i\eta^{\mu\nu},  \  \  \   \  \  (\eta^{\mu\nu}={\rm
diag}\{+1,-1,-1,-1\}), \lab{reco}
\eeq

\noindent
adopt the well known expressions, 

\begin{eqnarray}
&\langle x^{\mu}\vert X^{\mu}\vert\Psi\rangle =
x^{\mu}\Psi(x^{\mu}),\nonumber \\        
& \\
&\langle x^{\mu}\vert P_{\mu}\vert\Psi\rangle = i\frac{\partial}{\partial
x^{\mu}}
\Psi(x^{\mu}). \nonumber \lab{xp}
\end{eqnarray}

\noindent
(In Eq.  (\ref{reco}) $-i$  was  chosen  to preserve the sign in the ordinary
relations for the spatial part.)

In the Schr\"odinger picture, 

\beq
\vert\Psi(x^{\mu},\tau)\vert ^2, \lab{prob}
\eeq

\noindent
represents the probability density for the  system  to  be  at the space-time
point $x^{\mu}$ at ``instant'' $\tau$.  The  wave  function  evolves  with  a
Schr\"odinger equation, 

\beq
-i\frac{d}{d\tau}\Psi(x^{\mu},\tau) = {\cal H}\Psi(x^{\mu},\tau), \lab{ecs}
\eeq

\noindent
where 

\beq
{\cal H}=\frac{1}{2}\eta^{\mu\nu}P_{\mu}P_{\nu} \lab{ham} 
\eeq

\noindent
is a super-Hamiltonian (free case) which plays the  role  of  a mass operator
\cite{ena}. 

The  super-Hamiltonian  as  well  as  position  and  momentum  operators  are
Hermitian in the inner product (\ref{pe}).

A stationary solution of Eq. (\ref{ecs}) is 

\beq
\Psi(x^{\mu},\tau) = e^{i{\cal H}\tau}\psi(x^{\mu}), \lab{soes}
\eeq

\noindent
where $\psi(x^{\mu})$ is a solution of a generalized Klein-Gordon equation, 

\beq
\frac{\partial}{\partial x^{\mu}}\frac{\partial}{\partial
x_{\mu}}\psi(x^{\mu}) + m^2\psi(x^{\mu}) = 0, \lab{kg}
\eeq

\noindent
where  the  mass  eigenvalue  $m^2$ is not restricted {\it a  priori}  to  be
positive.    Thus,  the  theory  will  describe  real  (on-shell) and virtual
(off-shell) particles in a more symmetric way, which could be appropriate for
the curved space-time case, where the notion of real and virtual particles is
relative.

We can see that  the  theory  developed  is  formally  identical  to ordinary
quantum mechanics.  Then, all  we  have  learned  from  this  theory  can  be
rewritten in the PTF. 
 
The PTF can be formally obtained,  via  a  canonical  quantization procedure,
from  an  indefinite mass classical theory.   It  corresponds  to  a  unitary
realization of a symmetry group which, at the  classical level, is given by a
subgroup of canonical transformations, whose generating function reads  

\beq 
G_{XPM} = -\frac{\epsilon_{\mu\nu}}{2}M^{\mu\nu} + a^{\mu}P_{\mu} +
b_{\mu}X^{\mu}, \lab{xpm}
\eeq

\noindent 
where $\epsilon_{\mu\nu}=-\epsilon_{\nu\mu}$, $a^{\mu}$,  and  $b_{\mu}$  are
real parameters.  (Homogeneous Lorentz transformations are represented by the
unitary                operator                ${\cal         L}        =\exp
\{-\frac{i}{2}\epsilon_{\mu\nu}M^{\mu\nu}\}$,  under the scalar product given
in (\ref{pe}).) 

The algebra of the generators of such  a  group,  $X^{\mu}$,  $P_{\mu}$,  and
$M^{\mu\nu} =  X^{\mu}P^{\nu}  -  X^{\nu}P^{\mu}+ \frac{1}{2}\sigma^{\mu\nu}$
($\sigma^{\mu\nu}$ is the spin tensor), is known as $XPM$ algebra \cite{jon}.
The irreducible representations of  the  $XPM$ group are constructed from the
direct product of a covariant  version  of  the Heisenberg-Weyl group and the
homogeneous Lorentz group.  In the spin 0 case discussed above, this enlarged
Poincar\'e group can be represented in the wave function space defined in Eq.
(\ref{fo}).

Let  us  now come back to the localization  problem.    Note  that  from  Eq.
(\ref{reco}) it is immediate that, in general, 

\beq
[{\cal H}, X^{\mu}] \not= 0, \lab{loc}
\eeq

\noindent
while  the commutator vanishes in the standard (on-shell) RQM,  showing  that
the Heisenberg algebra (\ref{reco}) is only realizable in an indefinite  mass
theory.    Moreover,  from  (\ref{loc})  we  can readily see that it  is  not
possible to  localize  definite mass states preserving the Lorentz invariance
of localization.   Position operators such as the Newton-Wigner one \cite{nw}
are recovered on-shell, but  now  their interpretation can be clarified (Ref.
\cite{role}, and see also Ref.  \cite{hor}). 

The localization problem lies in the fact that the temporal coordinate has an
ambiguous character in RQM.  Such  a  problem  could be actually called ``the
problem of time in RQM''---at least for  the  usual  notion  of  time.   As a
consequence it is not possible, in this scheme,  to find a consistent quantum
mechanical  formulation for the relativistic particle (i.e., the construction
of a Hilbert space with a conserved non-negative norm).   This  issue  can be
bypassed  in two alternative ways.  One can consider the temporal  coordinate
as  a  parameter,  but then the one-particle interpretation fails.  Thus, one
can abandon  this  avenue  and turn to quantum field theory, where space-time
coordinates are parameters.    This  way  does  not  solve  the  localization
problem;  the problem  was swept under the rug \cite{ws}.  On the other hand,
using the PTF one can  return to a one-particle theory (completely equivalent
to quantum field theory but with  the significant advantage that one can deal
with a ``one-particle'' configuration space;  Ref.  \cite{ste}).  Introducing
an  evolution absolute parameter $\tau$, the space-time coordinates  can  now
acquire the same status as operators. 

PTF  ``solves''  the  problem  of  time in RQM.   However,  we  have  learned
something else:  the way to construct a consistent quantum  formalism  in  an
originally  reparametrization  invariant theory.\footnote{The usual  on-shell
covariant    action    for  a  spinless  particle    is    invariant    under
reparametrizations of the ``label time'' (which is  not  the parameter $\tau$
of our proposal).} This is the important analogy that we can mimic in QG.  Of
course, there is no problem with the role played  in  QG by time if we do not
adopt  the  canonical   quantization  procedure  \cite{dw2}.\footnote{In  the
context of minisuperspace models  one  can  overcome  the  problem of time by
adopting a ``third'' quantization theory.    See e.g., Ref.  \cite{homo}.} If
we adopt this proceduce, the problem  then  is that time is completely absent
in the Wheeler-DeWitt equation.  What is  still  unsolved is the construction
of a suitable quantum mechanical theory for a  wave  function  independent of
time.  The PTF provides a way. 

Let us compare the relativistic particle with the canonical formulation of QG
for minisuperspace models. 

The  analog  of  the  super-Hamiltonian  (\ref{ham})  is given (in the  $3+1$
decomposition of the metric) by

\beq
{\cal H}_{\rm m} = \frac{1}{2M^2}f^{ab}(q)p_a p_b + M^2 V(q) ,\lab{hwd}
\eeq

\noindent
which resembles the Hamiltonian of a particle of ``mass'' $M^2$ in  a  curved
background.    $f_{ab}$  is a metric on minisuperspace and $M$ is the  Planck
mass (the subscript  m  denotes  minisuperspace).    $p_a$  is  the  momentum
conjugate to $q^a$ which  represents  certain  components of the three-metric
and certain modes of the  matter  fields.    The term $V(q)$, which plays the
role of a potential, could also  be  introduced for the relativistic particle
but it does not modify the line of this discussion. 

We  can  establish  a  paralellism  with  the    formalism  developed  above.
Considering the configuration variables $q^a$, we can repeat  the  reasonings
of  Eqs.  (\ref{pe})--(\ref{prob}), defining the inner product (conserved  in
$\tau$) as  

\beq
\langle\Phi\vert\Psi\rangle = \int d^n q \sqrt{-f}\Phi^*(q^a)\Psi(q^a),
\lab{peqg}
\eeq

\noindent
where $n$ is the number of minisuperspace coordinates, and identifying  

\beq
X^{\mu} \longrightarrow \ q^a, \ \ \ \ \ P_{\mu} \longrightarrow \ p_a,
\lab{analog}
\eeq

\noindent
as canonical variables.  Thus, we can give a Hilbert  space  structure to the
space  of  wave  functions  $\Psi(q^a)$  by  demanding  that  they  be square
integrable with respect to $q^a$.  The corresponding evolution equation reads  

\beq
-i\frac{d}{d\tau}\Psi(q^a,\tau) = {\cal H}_{\rm m} \Psi(q^a,\tau), \lab{ecsqg}
\eeq

\noindent
where  ${\cal  H}_{\rm m}$ is given by Eq.  (\ref{hwd}).  (The time-dependent
Wheeler-DeWitt  equation    has    also  been  discussed  recently  in  Refs.
\cite{ht,uw}.) 

The  rest of  the  formalism  follows  the  outlines  of  quantum  mechanics.
Obviously, in this proposal  we  are  not  forced  to  restrict  the  allowed
``observables'' in the theory to  be those which commute with the Hamiltonian
(as  it  happens  in  the  ``naive  interpretation''  of  the  Wheeler-DeWitt
equation, Ref.  \cite{uw}).  Nevertheless, as  in  Eq.    (\ref{kg}),  we can
restrict the theory to admit only the on-shell  solutions  of  the eigenvalue
$0$, which correspond to solutions of the Wheeler-DeWitt equation.

The  retarded  propagator  of the Schr\"odinger equation (\ref{ecsqg}), which
gives  the transition amplitude for the system ``to evolve'' between  $q_0^a$
at  $\tau=0$  and  $q^{a}$ at $\tau$, is just \ $\theta(\tau)\langle q^a\vert
e^{i{\cal H}_{\rm m} \tau}\vert  q_0^{a}\rangle$.    The  corresponding  time
independent Green function reads  

\beq
G(q^a, q_0^{a}) = \int_{0}^{+\infty}d\tau \langle q^a\vert
e^{i{\cal H}_{\rm m}\tau}\vert q_0^a\rangle. \lab{ges}
\eeq

\noindent
In  the  analogous expression for  the  relativistic  particle,  selecting  a
particular value of the mass and  putting  a regulator $+i0$ term in order to
damp  the  oscillations  for large $\tau$, one  obtains  the  standard  Green
function  for  the  Klein-Gordon equation, with the Feynman  prescription  to
avoid  the  poles.    This  is  just  the result  obtained  from  the  second
quantization of the Klein-Gordon field.  In the same way  the  Green function
(\ref{ges})  is  equal  to  those derived by proceeding to the quantum  field
theory of the Universe (``third quantization''). 

The  propagator in Eq.  (\ref{ges}) can be conveniently expressed via a  path
integral.   The  derivation essentially follows the same steps as in ordinary
quantum mechanics (see,  e.g.,  Ref.  \cite{ram};  the original derivation of
the path integral of  a  particle  in  curved space-time can be found in Ref.
\cite{orig}).  It gives  

\beq
G(q^a, q_0^{a}) = \int_0^{\infty} d\tau \int {\cal D}q {\cal D}p \exp
\{-i\int_0^{\tau}[p_a \dot q^a - {\cal H}_{\rm m}]d\tau'\}, \lab{green}
\eeq

\noindent
where we can see the classical action $S = \int_0^{\tau}[p_a \dot q^a - {\cal
H}_{\rm m}]d\tau'$ corresponding to the ``free  constraint''  theory.   These
calculations are examples of a very important  property of the PTF.  Since in
the theory, the constraint ${\cal H}_{\rm m} =  0$  is  not present, the path
integral does not present any difficulty, unlike the constrained system.  The
same  result (\ref{green}) has been recently obtained by Halliwell \cite{hal}
using  the  method  introduced by Batalin, Fradkin, and Vilkovisky \cite{bfv}
based on  the  BRST  invariance.    To  compare  his result with ours we must
identify $d\tau$ with  $dN t$, where $N$ is the lapse function appearing as a
Lagrange multiplier in the action $S = \int_0^t[p_a \dot q^a - N{\cal H}_{\rm
m}]dt'$.    The  latter identification  was  originally  made  by  Teitelboim
\cite{tei} for the relativistic particle as  well as the gravitational field,
imposing a gauge fixing condition.  Even if the different approaches give the
same expression for the Feynman Green function, we  believe that our proposal
is not only a very simple way to attack the problem of time but also gains in
physical  content.\footnote{As  was  pointed out by Stephens \cite{ste} there
are several  different  approaches by which one may arrive at the PTF for the
relativistic particle and  the  same  statement  is valid for QG.  From these
alternatives we have chosen  the Feynman proposal which has a richer physical
content than the others.}    

The original goal of this  essay  was  to establish a paralellism between the
localization and the time problems.   However, a door is open:  the off-shell
theory (such as the virtual particles appearing  in  the  PTF  of  RQM).  Its
physical implications and a more detailed discussion will  be considered in a
separate work \cite{cgga}. 

\vspace{0.5in}
\section*{ACKNOWLEDGMENTS}

This essay  received  an  Honorable  Mention in the 1994 Awards for Essays on
Gravitation  of  the    Gravity  Research  Foundation,  which  we  gratefully
acknowledge. 

We  would like to thank Claudia Bernardou, Luca Bombelli,  Mario  Castagnino,
Graciela  Garcia Alvarez, and Sebastiano Sonego for their help.   One  of  us
(F.H.G.) wishes to thank Edgard Gunzig and Ilya Prigogine for hospitality  at
the Universit\'e Libre de Bruxelles when this work was carried out.

\end{document}